\documentclass[twocolumn,preprintnumbers,showpacs,amsmath,amssymb]{revtex4}
\usepackage{docs}

\usepackage{graphicx}
\usepackage{dcolumn}

\usepackage{bm}

\begin{document}
\title{Stable dark energy stars}%

\author{Francisco S. N. Lobo}%
\email{flobo@cosmo.fis.fc.ul.pt}
\affiliation{Centro de Astronomia
e Astrof\'{\i}sica da Universidade de Lisboa,\\
Campo Grande, Ed. C8 1749-016 Lisboa, Portugal}

\begin{abstract}

The gravastar picture is an alternative model to the concept of a
black hole, where there is an effective phase transition at or
near where the event horizon is expected to form, and the interior
is replaced by a de Sitter condensate. In this work, a
generalization of the gravastar picture is explored, by
considering a matching of an interior solution governed by the
dark energy equation of state, $\omega\equiv p/ \rho<-1/3$, to an
exterior Schwarzschild vacuum solution at a junction interface.
The motivation for implementing this generalization arises from
the fact that recent observations have confirmed an accelerated
cosmic expansion, for which dark energy is a possible candidate.
Several relativistic dark energy stellar configurations are
analyzed by imposing specific choices for the mass function. The
first case considered is that of a constant energy density, and
the second choice, that of a monotonic decreasing energy density
in the star's interior. The dynamical stability of the transition
layer of these dark energy stars to linearized spherically
symmetric radial perturbations about static equilibrium solutions
is also explored. It is found that large stability regions exist
that are sufficiently close to where the event horizon is expected
to form, so that it would be difficult to distinguish the exterior
geometry of the dark energy stars, analyzed in this work, from an
astrophysical black hole.

\end{abstract}
\pacs{04.20.Jb, 04.40.Dg, 97.10.-q}
\maketitle

\def\d{{\mathrm{d}}}

\section{Introduction}

The structure of relativistic stars and the phenomenon of
gravitational collapse is of fundamental importance in
astrophysics and has attracted much attention in the relativist
community since the formulation of general relativity. Relatively
to the construction of theoretical models of relativistic stars,
one may refer to the pioneering work of Schwarzschild \cite{Schw},
Tolman \cite{Tolman}, and Oppenheimer and Volkoff \cite{OppVolk}.
Schwarzschild \cite{Schw} considered analytic solutions describing
a star of uniform density, and Tolman \cite{Tolman} developed a
method providing explicit solutions for static spheres of fluid,
which proved important to the study of stellar structure.
Oppenheimer and Volkoff \cite{OppVolk} by suitably choosing
specific Tolman solutions studied the gravitational equilibrium of
neutron stars, using the equation of state for a cold Fermi gas,
consequently laying down the foundations of the general
relativistic theory of stellar structures. Paging through history
one finds an extremely extensive literature~\cite{Gravitation},
however, one may refer to the important contributions of
Chandrasekhar \cite{Chand} in the construction of white-dwarf
models by taking into account special relativistic effects in the
electron degeneracy equation of state, where he discovered that no
white dwarf may have a mass greater than $\sim 1.2$ solar masses,
which has been called the Chandrasekhar limit; and the work of
Baade and Zwicky \cite{neutron}, where they invented the concept
of a neutron star, and identified astronomical objects denoted as
supernovae, representing a transitional collapse of an ordinary
star into a neutron star. Far from undertaking an exhaustive
review, one may also refer to the work of Wyman, where in Ref.
\cite{Wyman1} isotropic coordinates were used to solve the
relativistic equations of a perfect fluid with constant energy
density, and in Ref. \cite{Wyman2}, a critical examination and
generalization of the Tolman solutions was undertaken; Buchdahl
\cite{Buchdahl} and Bondi \cite{Bondi} generalized the interior
Schwarzschild solution to more general static fluid spheres in the
form of inequalities involving the mass concentration, central
energy density and central pressure; Leibovitz \cite{Leib} also
generalized some of Tolman's solutions by applying a more physical
approach; and the discovery of pulsars by Hewish {\it et al}
\cite{Hewish}, the idea advanced by Gold \cite{Gold} that pulsars
might be rotating neutron stars, which was later confirmed by
observations, and the idealized models of rotating neutron stars
\cite{rotate-neutron}. It is also interesting to note that Bayin
found new solutions for static fluid spheres \cite{Bayin1},
explored the time-dependent field equations for radiating fluid
spheres \cite{Bayin2}, and further generalized the analysis to
anisotropic fluid spheres \cite{Bayin3}.

Relatively to the issue of gravitational collapse, before the
mid-1960s, the object now known as a black hole, was referred to
as a collapsed star \cite{Thorne-etal}. Oppenheimer and Snyder
\cite{OppSnyd}, in 1939, provided the first insights of the
gravitational collapse into a black hole, however, it was only in
1965 that marked an era of intensive research into black hole
physics. Although evidence for the existence of black holes is
very convincing, it has recently been argued that the
observational data can provide very strong arguments in favor of
the existence of event horizons but cannot fundamentally prove it
\cite{AKL}. This scepticism has inspired new and fascinating
ideas. In this line of thought, it is interesting to note that a
new final state of gravitational collapse has been proposed by
Mazur and Mottola \cite{Mazur}. In this model, and in the related
picture developed by Laughlin {\it et al} \cite{Laughlin}, the
quantum vacuum undergoes a phase transition at or near the
location where the event horizon is expected to form. The model
denoted as a gravastar ({\it grav}itational {\it va}cuum {\it
star}), consists of a compact object with an interior de Sitter
condensate, governed by an equation of state given by $p=-\rho$,
matched to a shell of finite thickness with an equation of state
$p=\rho$. The latter is then matched to an exterior Schwarzschild
vacuum solution. The thick shell replaces both the de Sitter and
the Schwarzschild horizons, therefore, this gravastar model has no
singularity at the origin and no event horizon, as its rigid
surface is located at a radius slightly greater than the
Schwarzschild radius. It has been argued that there is no way of
distinguishing a Schwarzschild black hole from a gravastar from
observational data \cite{AKL}. It was also further shown by Mazur
and Mottola that gravastars are thermodynamically stable. Related
models, analyzed in a different context have also been considered
by Dymnikova \cite{Dymnikova}.
However, in a simplified model of the Mazur-Mottola picture,
Visser and Wiltshire \cite{VW} constructed a model by matching an
interior solution with $p=-\rho$ to an exterior Schwarzschild
solution at a junction interface, comprising of a thin shell. The
dynamic stability was then analyzed, and it was found that some
physically reasonable stable equations of state for the transition
layer exist. In Ref. \cite{CFV}, a generalized class of similar
gravastar models that exhibit a continuous pressure profile,
without the presence of thin shells was analyzed. It was found
that the presence of anisotropic pressures are unavoidable, and
the TOV equation was used to place constraints on the anisotropic
parameter. It was also found that the transverse pressures permit
a higher compactness than is given by the Buchdahl-Bondi bound
\cite{Buchdahl,Bondi} for perfect fluid stars, and several
features of the anisotropic equation of state were explored. In
Ref. \cite{Bilic}, motivated by low energy string theory, an
alternative model was constructed by replacing the de Sitter
regime with an interior solution governed by a Chaplygin gas
equation of state, interpreted as a Born-Infeld phantom gravastar.

It has also been recently proposed by Chapline that this new
emerging picture consisting of a compact object resembling
ordinary spacetime, in which the vacuum energy is much larger than
the cosmological vacuum energy, has been denoted as a ``dark
energy star'' \cite{Chapline}. In fact, in the present paper a
mathematical model generalizing the Mazur-Mottola picture, or for
that matter, the Visser-Wiltshire model is proposed, where the
interior de Sitter solution is replaced by a solution governed by
the dark energy equation of state, $p=\omega \rho$ with
$\omega<-1/3$, matched to an exterior vacuum Schwarzschild
solution. Note that the particular case of $\omega=-1$ reduces to
the Visser-Wiltshire gravastar model.
The motivation for implementing this generalization comes from the
fact that recent observations have confirmed that the Universe is
undergoing a phase of accelerated expansion. Evidence of this
cosmological expansion has been shown independently from
measurements of supernovae of type Ia (SNe Ia) \cite{supernovae}
and from cosmic microwave background radiation \cite{CMB}. A
possible candidate proposed for this cosmic acceleration is
precisely that of dark energy, a cosmic fluid parametrized by an
equation of state $\omega \equiv p/\rho<-1/3$, where $p$ is the
spatially homogeneous pressure and $\rho$ the dark energy density.
If $\omega<-1$, a case certainly not excluded, and in fact
favored, by observations, the null energy condition is violated.
For this case, the cosmic fluid is denoted phantom energy, and
possesses peculiar properties, such as negative temperatures
\cite{dark-entropy} and the energy density increases to infinity
in a finite time, resulting in a Big Rip \cite{Weinberg}. It also
provides one with a natural scenario for the existence of exotic
geometries such as wormholes \cite{Sushkov,Lobo-phantom}. In this
context, it was also shown that the masses of all black holes tend
to zero as the phantom energy universe approaches the Big
Rip~\cite{Babichev}. However, it is interesting that recent fits
to supernovae, CMB and weak gravitational lensing data favor an
equation of state with a dark energy parameter crossing the
phantom divide $\omega=-1$ \cite{Vikman,phantom-divide}. In a
cosmological setting the transition into the phantom regime, for a
single scalar field \cite{Vikman} is probably physically
implausible, thus the stress energy tensor should include a
mixture of various interacting non-ideal fluids.

As emphasized in Refs. \cite{Sushkov,Lobo-phantom}, in a rather
different context, a subtlety needs to be pointed out: The notion
of dark energy is that of a spatially homogeneous cosmic fluid,
however, it can be extended to inhomogeneous spherically symmetric
spacetimes by regarding that the pressure in the dark energy
equation of state is a negative radial pressure, and the
transverse pressure may be determined via the field equations. (An
inhomogeneous spherically symmetric dark energy scalar field was
also considered in Ref. \cite{Picon-Lim}). In this context, the
generalization of the gravastar picture with the inclusion of an
interior solution governed by the equation of state, $p=\omega
\rho$ with $\omega<-1/3$, will be denoted by a dark energy
gravastar, or simply a ``dark energy star'' in agreement with the
Chapline definition \cite{Chapline}. We shall explore several
configurations, by imposing specific choices for the mass
function. We shall then explore the dynamical stability of the
transition layer of these models to linearized perturbations
around static solutions, by applying the general stability
formalism developed in Ref. \cite{LoboCrawford}, and which was
also applied in the context of the stability of phantom wormholes
\cite{stableWH}, and further analyze the evolution identity to
extract some physical insight regarding the pressure balance
equation across the junction interface.
The dark energy star outlined in this paper may possibly have an
origin in a density fluctuation in the cosmological background. It
is uncertain how such inhomogeneities in the dark energy may be
formed. However, a possible explanation may be inferred from Ref.
\cite{inhomogEOS}, where the dark energy equation of state was
generalized to include an inhomogeneous Hubble parameter dependent
term, possibly resulting in the nucleation of a dark energy star
through a density perturbation.

This paper is outlined in the following manner. In Section II, we
present the structure equations of dark energy stellar models. In
Section III, specific models are then analyzed by imposing
particular choices for the mass function. In Section IV, the
linearized stability analysis procedure is briefly outlined, and
the stability regions of the transition layer of specific dark
energy stars are determined. Finally in Section V, we conclude.

\section{Dark energy stars: Equations of structure}


Consider the interior spacetime, without a loss of generality,
given by the following metric, in curvature coordinates
\begin{eqnarray}
ds^2&=&-{\rm exp}\left[-2\int_r^{\infty}
g(\tilde{r})d\tilde{r}\right]\,dt^2+\frac{dr^2}{1- 2m(r)/r}
    \nonumber   \\
&&+r^2 \,(d\theta ^2+\sin ^2{\theta} \, d\phi ^2) \label{metric}
\,,
\end{eqnarray}
where $g(r)$ and $m(r)$ are arbitrary functions of the radial
coordinate, $r$. The function $m(r)$ is the quasi-local mass, and
is denoted as the mass function. The factor $g(r)$ is the
``gravity profile'' and is related to the locally measured
acceleration due to gravity, through the following relationship:
${\cal A}=\sqrt{1-2m(r)/r}\,g(r)$ \cite{Martin,Martin2}. The
convention used is that $g(r)$ is positive for an inwardly
gravitational attraction, and negative for an outward
gravitational repulsion. Note that equivalently one may consider a
function $\Phi(r)$, defined as $\Phi(r)=-\int_r^{\infty}
g(\tilde{r})d\tilde{r}$, and denoted as the redshift function, as
it is related to the gravitational redshift \cite{Morris}.


The stress-energy tensor for an anisotropic distribution of matter
is provided by
\begin{equation}
T_{\mu\nu}=(\rho+p_t)U_\mu \, U_\nu+p_t\,
g_{\mu\nu}+(p_r-p_t)\chi_\mu \chi_\nu \,,
\end{equation}
where $U^\mu$ is the four-velocity, $\chi^\mu$ is the unit
spacelike vector in the radial direction, i.e.,
$\chi^\mu=\sqrt{1-2m/r}\,\delta^\mu{}_r$. $\rho(r)$ is the energy
density, $p_r(r)$ is the radial pressure measured in the direction
of $\chi^\mu$, and $p_t(r)$ is the transverse pressure measured in
the orthogonal direction to $\chi^\mu$.

Thus, the Einstein field equation, $G_{\mu\nu}=8\pi T_{\mu\nu}$,
where $G_{\mu\nu}$ is the Einstein tensor, provides the following
relationships
\begin{eqnarray}
m'&=&4\pi r^2 \rho   \label{mass}\,,\\
g&=&\frac{m+4\pi r^3 p_r}{r(r-2m)} \label{Phi}\,,\\
p_r'&=&-\frac{(\rho+p_r)(m+4\pi r^3 p_r)}{r(r-2m)}
+\frac{2}{r}(p_t-p_r)\label{anisotTOV}\,.,
\end{eqnarray}
where the prime denotes a derivative with respect to the radial
coordinate, $r$. Equation (\ref{anisotTOV}) corresponds to the
anisotropic pressure Tolman-Oppenheimer-Volkoff (TOV) equation.

Now, using the dark energy equation of state, $p_r=\omega \rho$,
and taking into account Eqs. (\ref{mass}) and (\ref{Phi}), we have
the following relationship
\begin{equation}
g(r)=\frac{m+\omega rm'}{r\,\left(r-2m \right)} \,.
            \label{EOScondition}
\end{equation}
There is, however, a subtle point that needs to be emphasized
\cite{Sushkov,Lobo-phantom}. The notion of dark energy is that of
a spatially homogeneous cosmic fluid. Nevertheless, it can be
extended to inhomogeneous spherically symmetric spacetimes, by
regarding that the pressure in the equation of state $p=\omega
\rho$ is a radial pressure, and that the transverse pressure may
be obtained from Eq. (\ref{anisotTOV}). Note that for the
particular case of $\omega=-1$, from Eq. (\ref{EOScondition}), one
has the following solution $g_{tt}=-(1-2m/r)$, which reduces to
the specific class of solutions analyzed in Ref. \cite{VW}.

Using the dark energy equation of state $p_r=\omega \rho$, Eq.
(\ref{anisotTOV}) in terms of the principal pressures, takes the
form
\begin{eqnarray}
p_r'=-p_r\left(\frac{1+\omega}{\omega}\right)\,\frac{m+\omega
rm'}{r\,\left(r-2m \right)} +\frac{2}{r}\,(p_t-p_r)
    \label{TOVdark}   \,,
\end{eqnarray}
which taking into account Eq. (\ref{mass}), may be expressed in
the following equivalent form
\begin{eqnarray}
\Delta=\frac{\omega}{8\pi r^2}\left[m''
r-2m'+\left(\frac{1+\omega}{\omega}\right)m'r g \right]
   \label{TOVdark2}   \,.
\end{eqnarray}
$\Delta=p_t-p_r$ is denoted as the anisotropy factor, as it is a
measure of the pressure anisotropy of the fluid comprising the
dark energy star. $\Delta=0$ corresponds to the particular case of
an isotropic pressure dark energy star. Note that $\Delta/r$
represents a force due to the anisotropic nature of the stellar
model, which is repulsive, i.e., being outward directed if
$p_t>p_r$, and attractive if $p_t<p_r$.

One now has at hand four equations, namely, the field equations
(\ref{mass})-(\ref{anisotTOV}) and Eq. (\ref{EOScondition}), with
five unknown functions of $r$, i.e., $\rho(r)$, $p_r(r)$,
$p_t(r)$, $g(r)$ and $m(r)$. Obtaining explicit solutions to the
Einstein field equations is extremely difficult due to the
nonlinearity of the equations, although the problem is
mathematically well-defined. However, in the spirit of Ref.
\cite{Lobo-phantom}, we shall adopt the approach in which a
specific choice for a physically reasonable mass function $m(r)$
is provided and through Eq. (\ref{EOScondition}), $g(r)$ is
determined, thus consequently providing explicit expressions for
the stress-energy tensor components.
In the specific cases that follow, we shall consider that the
energy density be positive and finite at all points in the
interior of the dark energy star.

\section{Specific models}

\subsection{Constant energy density}

Consider the specific case of a constant energy density,
$\rho(r)=\rho_0$, so that Eq. (\ref{mass}) provides the following
mass function
\begin{equation}
m(r)=\frac{4\pi \rho_0}{3} r^3 \,.
      \label{const-rho}
\end{equation}
Thus, using Eq. (\ref{EOScondition}), one finds that $g(r)$ is
given by
\begin{equation}
g(r)=\frac{Ar(1+3\omega)}{1-2Ar^2} \,,
\end{equation}
where for simplicity, the definition $A=4\pi \rho_0/3$ is used.
Note that for $\omega<-1/3$, we have an outward gravitational
repulsion, $g(r)<0$, which is to be expected in gravastar models.

The spacetime metric for this solution takes the following form
\begin{eqnarray}
ds^2&=&-\left(1-2Ar^2\right)^{-(1+3\omega)/2}\,dt^2+\frac{dr^2}{1-2Ar^2}
    \nonumber   \\
&&+r^2 \,(d\theta ^2+\sin ^2{\theta} \, d\phi ^2)   \,.
\end{eqnarray}
The stress-energy tensor components are given by $p_r=\omega
\rho_0$ and
\begin{equation}
p_t=\omega\rho_0\left[1+\frac{(1+\omega)(1+3\omega)Ar^2}{2\omega(1-2Ar^2)}\right]
\,.
\end{equation}

\begin{figure}[h]
\centering
  \includegraphics[width=2.8in]{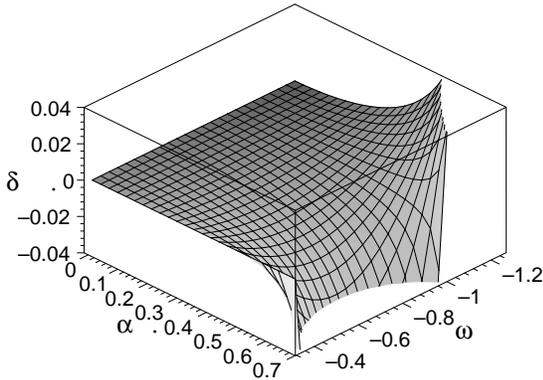}
  \caption{Plot of the anisotropy factor for a dark energy star
  with a constant energy density. We have defined the following
  dimensionless parameters: $\delta=\Delta/A$ and $\alpha=\sqrt{A}r$.
  We verify that $\Delta<0$ in the range $-1/3>\omega>-1$ and
  $\Delta>0$ in the phantom regime, $\omega<-1$.}
  \label{Fig:Delta1}
\end{figure}

The anisotropy factor is provided by
\begin{equation}
\Delta=\frac{3}{8\pi} (1+\omega)(1+3\omega)\frac{A^2r^2}{1-2Ar^2}
\,.
\end{equation}
One readily verifies that $\Delta<0$ for $-1<\omega<-1/3$ and
$\Delta>0$ in the phantom regime, $\omega<-1$. However, it is
perhaps instructive to plot $\Delta$, which is depicted in Fig.
\ref{Fig:Delta1}. Note that $\Delta=0$ at the origin, $r=0$, as
was to be expected. For $\omega=-1$, then $\Delta=0$ is also
verified for arbitrary $r$. This latter condition is also readily
derived from Eq. (\ref{anisotTOV}), where taking into account
$p_r=\omega\rho$ for $p_r={\rm const}$, one verifies $p_t=p_r$.

\subsection{Tolman-Matese-Whitman mass function}

Consider the following choice for the mass function, given by
\begin{equation}
m(r)=\frac{b_0 r^3}{2(1+2b_0 r^2)}\,,
      \label{TMWmass}
\end{equation}
where $b_0$ is a non-negative constant. The latter may be
determined from the regularity conditions and the finite character
of the energy density at the origin $r=0$, and is given by
$b_0=8\pi \rho_c/3$, where $\rho_c$ is the energy density at
$r=0$.

This choice of the mass function represents a monotonic decreasing
energy density in the star interior, and was used previously in
the analysis of isotropic fluid spheres by Matese and Whitman
\cite{MatWhit} as a specific case of the Tolman type$-IV$ solution
\cite{Tolman}, and later by Finch and Skea~\cite{Finch}.
Anisotropic stellar models, with the respective astrophysical
applications, were also extensively analyzed in Refs. \cite{Mak},
by considering a specific case of the Matese-Whitman mass
function. The numerical results outlined show that the basic
physical parameters, such as the mass and radius, of the model can
describe realistic astrophysical objects such as neutron stars
\cite{Mak}.

\begin{figure}[h]
\centering
  \includegraphics[width=2.8in]{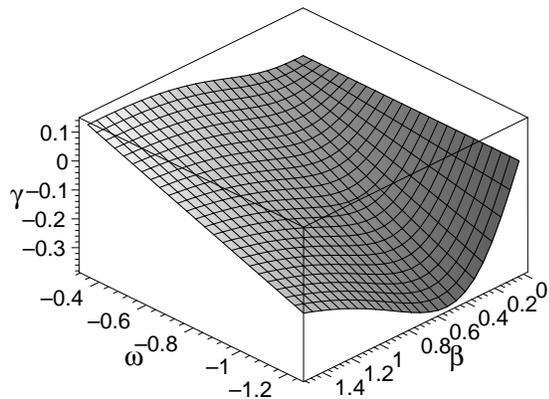}
  \caption{Plot of the ``gravity profile'' for the
  Tolman-Matese-Whitman mass function. We have defined the
  following dimensionless parameters: $\gamma=g(r)/\sqrt{b_0}$ and
  $\beta=\sqrt{b_0}r$. Note that, qualitatively, $g(r)$ takes
  positive values of $\omega$ in the neighborhood of $-1/3$,
  indicating an inwardly gravitational attraction. To be a solution
  of a gravastar, it is necessary that the local acceleration due
  to gravity of the interior solution be repulsive, so that the
  region for which $g(r)>0$ is necessarily excluded.
  See the text for details.}
  \label{Fig:g2}
\end{figure}

Using Eq. (\ref{EOScondition}), $g(r)$ is given by
\begin{equation}
g(r)=\left(\frac{b_0r}{2}\right)\left[\frac{(1+3\omega)
+(1+\omega)2b_0r^2}{(1+b_0r^2)(1+2b_0r^2)}\right] \,,
\end{equation}
which is plotted in Fig. \ref{Fig:g2}. Note that $g(r)>0$, for
$\omega>-(1+2b_0r^2)/(3+2b_0r^2)$, indicating an inwardly
gravitational attraction; from the plot one verifies that,
qualitatively, $g(r)$ is positive for values of $\omega$ in the
neighborhood of $-1/3$.
Now, to be a solution of a gravastar, it is necessary that the
local acceleration due to gravity of the interior solution be
repulsive, so that the region above-mentioned for which $g(r)>0$
is necessarily excluded.

The spacetime metric for this solution is provided by
\begin{eqnarray}
ds^2&=&-\left(1+b_0r^2\right)^{(1-\omega)/2}
\left(1+2b_0r^2\right)^{\omega}\,dt^2
    \nonumber   \\
&&\hspace{-0.5cm}+\left(\frac{1+2b_0r^2}{1+b_0r^2}\right)dr^2+r^2
\,(d\theta ^2+\sin ^2{\theta} \, d\phi ^2)   \,.
\end{eqnarray}
The stress-energy tensor components are given by
\begin{eqnarray}
p_r&=&\omega \rho =\left( \frac{\omega
b_0}{8\pi}\right)\left[\frac{3+2b_0 r^2}{(1+2b_0 r^2)^2} \right]
    \nonumber   \\
p_t&=& -\left( \frac{b_0}{8\pi}\right)\left[\frac{\omega(3+2b_0
r^2)}{(1+2b_0 r^2)^2}\right]+\left(\frac{b_0^2r^2}{32\pi}
\right)\times
     \nonumber   \\
&&\hspace{-0.5cm}\times\Big\{(1+\omega)(3+2b_0
r^2)\left[(1+3\omega)+2b_0r^2(1+\omega)\right]
    \nonumber   \\
&&\hspace{-1.25cm}-8\omega(5+2b_0r^2)(1+b_0r^2)\Big\} \big/
\left[(1+2b_0 r^2)^3(1+b_0 r^2)\right]   .
\end{eqnarray}

\begin{figure}[h]
\centering
  \includegraphics[width=2.8in]{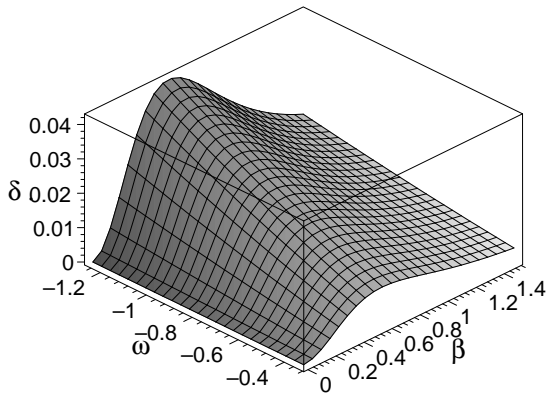}
  \caption{Plot of the anisotropy factor for the
  Tolman-Matese-Whitman mass function. We have defined the following
  dimensionless parameters: $\delta=\Delta/b_0$ and
  $\beta=\sqrt{b_0}r$. Note that $\Delta>0$ for generic $\omega$ and
  $r$.}
  \label{Fig:Delta2}
\end{figure}

The anisotropy factor takes the following form
\begin{eqnarray}
\Delta&=&\frac{b_0^2r^2}{32\pi}\Big\{(1+\omega)(3+2b_0
r^2)\left[(1+3\omega)+2b_0r^2(1+\omega)\right]
    \nonumber   \\
&&\hspace{-0.98cm}-8\omega(5+2b_0r^2)(1+b_0r^2)\Big\}\big/\left[(1+2b_0
r^2)^3(1+b_0 r^2)\right] ,
\end{eqnarray}
which is plotted in Fig. \ref{Fig:Delta2}, where one verifies
$\Delta>0$. For the particular case of $\omega=-1$, the anisotropy
factor reduces to
\begin{equation}
\Delta=\left(\frac{b_0^2r^2}{4\pi}\right)\frac{(5+2b_0
r^2)}{(1+2b_0 r^2)^3}\,.
\end{equation}

\section{Stability of dark energy stars}

\subsection{Junction conditions}

In this article, we shall model dark energy stars by matching an
interior solution, governed by an equation of state, $p=\omega
\rho$ with $\omega<-1/3$, to an exterior Schwarzschild vacuum
solution with $p=\rho=0$, at a junction interface $\Sigma$, with
junction radius $a$. The Schwarzschild metric is given by
\begin{equation}
ds^2=-\left(1-\frac{2M}{r}\right)\,dt^2+\frac{dr^2}{1-
\frac{2M}{r}}+r^2 \,(d\theta ^2+\sin ^2{\theta} \, d\phi ^2)
\label{Sch-metric}\,,
\end{equation}
which possesses an event horizon at $r_b=2M$. To avoid the latter,
the junction radius lies outside $2M$, i.e., $a>2M$. We shall show
below that $M$, in this context, may be interpreted as the total
mass of the dark energy star. (In an analogous manner traversable
wormhole models were constructed in Refs. \cite{wormhole-Lobo} and
the stability analysis of phantom wormholes was carried out in
Ref. \cite{stableWH}.) One may also impose boundary and regularity
conditions, and their respective constraints
\cite{Martin,bound-constraints}, at the center and at the surface
of the dark energy star, without the presence of thin shells,
however, this analysis will be presented in a subsequent paper.

Using the Darmois-Israel formalism~\cite{Darmois-Israel}, the
surface stress-energy tensor $S^i_j$ at the junction surface
$\Sigma$ is given by the Lanczos equations,
$S^i_j=-\frac{1}{8\pi}(\kappa^i_j-\delta^i_j \kappa^k_k)$.
$\kappa^i_j$ is defined by the discontinuity of the extrinsic
curvatures, $K^i_j$, across the junction interface, i.e.,
$\kappa_{ij}=K^+_{ij}-K^-_{ij}$.
The extrinsic curvature is defined as
$K_{ij}=n_{\mu;\nu}\,e^{\mu}_{(i)}e^{\nu}_{(j)}$, where $n^{\mu}$
is the unit normal $4-$vector to $\Sigma$, and $e^{\mu}_{(i)}$ are
the components of the holonomic basis vectors tangent to $\Sigma$.
Using the metrics (\ref{metric}) and (\ref{Sch-metric}), the
non-trivial components of the extrinsic curvature are given by
(see Ref. \cite{LoboCrawford} for further details)
\begin{eqnarray}
K ^{\tau
\;+}_{\;\;\tau}&=&\frac{\frac{M}{a^2}+\ddot{a}}{\sqrt{1-\frac{2M}{a}+\dot{a}^2}}
\;,  \label{Kplustautau2}\\
K ^{\tau \;-}_{\;\;\tau}&=&\frac{\frac{m+\omega
am'}{a^2}+\ddot{a}-\frac{(1+\omega)m'\dot{a}^2}{(a-2m)}}{\sqrt{1-\frac{2m(a)}{a}+\dot{a}^2}}
\;, \label{Kminustautau2}
\end{eqnarray}
and
\begin{eqnarray}
K ^{\theta
\;+}_{\;\;\theta}&=&\frac{1}{a}\sqrt{1-\frac{2M}{a}+\dot{a}^2}\;,
 \label{Kplustheta2}\\
K ^{\theta
\;-}_{\;\;\theta}&=&\frac{1}{a}\sqrt{1-\frac{2m(a)}{a}+\dot{a}^2}
\;,  \label{Kminustheta2}
\end{eqnarray}
where the overdot denotes a derivative with respect to the proper
time, $\tau$, and the prime henceforth shall denote a derivative
with respect to $a$. Equation (\ref{EOScondition}) evaluated at
$a$ has been used to eliminate $g(r)$ from Eq.
(\ref{Kminustautau2}). In the analysis that follows, we shall
apply the same procedure, so that in deducing the master equation,
which dictates the stability regions, one shall only make use of
the mass function $m(r)$.
Thus, the Lanczos equations provide the surface stresses given by
\begin{eqnarray}
\sigma&=&-\frac{1}{4\pi a} \left(\sqrt{1-\frac{2M}{a}+\dot{a}^2}-
\sqrt{1-\frac{2m}{a}+\dot{a}^2} \, \right)
    \label{surfenergy}   ,\\
{\cal P}&=&\frac{1}{8\pi a} \Bigg(\frac{1-\frac{M}{a}
+\dot{a}^2+a\ddot{a}}{\sqrt{1-\frac{2M}{a}+\dot{a}^2}}
      \nonumber    \\
&&- \frac{1+\omega
m'-\frac{m}{a}+\dot{a}^2+a\ddot{a}+\frac{\dot{a}^2m'(1+\omega)}{1-2m/a}}
{\sqrt{1-\frac{2m}{a}+\dot{a}^2}} \, \Bigg)
    \label{surfpressure}    \,,
\end{eqnarray}
$\sigma$ and ${\cal P}$ are the surface energy density and the
tangential surface pressure, respectively.

We shall also use the conservation identity given by
$S^{i}_{j|i}=\left[T_{\mu \nu}e^{\mu}_{(j)}n^{\nu}\right]^+_-$,
where $[X]^+_-$ denotes the discontinuity across the surface
interface, i.e., $[X]^+_-=X^+|_{\Sigma}-X^-|_{\Sigma}$. The
momentum flux term in the right hand side corresponds to the net
discontinuity in the momentum flux $F_\mu=T_{\mu\nu}\,U^\nu$ which
impinges on the shell. The conservation identity is a statement
that all energy and momentum that plunges into the thin shell,
gets caught by the latter and converts into conserved energy and
momentum of the surface stresses of the junction.

Note that $S^{i}_{\tau|i}=-\left[\dot{\sigma}+2\dot{a}(\sigma
+{\cal P} )/a \right]$, and the flux term is given by (see Ref.
\cite{LoboCrawford} for details)
\begin{equation}
\left[T_{\mu
\nu}e^{\mu}_{(j)}n^{\nu}\right]^+_-=-\frac{(\rho+p_r)\dot{a}
\sqrt{1-2m/a+\dot{a}^2}}{1-2m/a} \,,
\end{equation}
where $\rho$ and $p_r$ may be deduced from Eqs.
(\ref{mass})-(\ref{Phi}), respectively, evaluated at the junction
radius, $a$.
Therefore, using these relationships, the conservation identity
provides us with
\begin{equation}
\sigma'=-\frac{2}{a}\,(\sigma+{\cal P})+\Xi
  \,,\label{consequation2}
\end{equation}
where $\Xi$, defined for notational convenience, is given by
\begin{eqnarray}
\Xi=-\frac{1}{4\pi a} \frac{m'(1+\omega)}{\left(a-2m \right)}
 \sqrt{1-\frac{2m}{a}+\dot{a}^2} \,,
  \label{flux-Xi}
\end{eqnarray}
using Eq. (\ref{EOScondition}) evaluated at $a$. Note that the
flux term is zero when $\omega=-1$, which reduces to the analysis
considered in Ref. \cite{VW}.

Taking into account Eqs. (\ref{surfenergy})-(\ref{surfpressure})
and Eq. (\ref{flux-Xi}), then Eq. (\ref{consequation2}) finally
takes the form
\begin{eqnarray}
\sigma'&=&\frac{1}{4\pi a^2} \Bigg(\frac{1-\frac{3M}{a}
+\dot{a}^2-a\ddot{a}}{\sqrt{1-\frac{2M}{a}+\dot{a}^2}}
\nonumber   \\
&& - \frac{1-\frac{3m}{a}+m'+\dot{a}^2
-a\ddot{a}}{\sqrt{1-\frac{2m}{a}+\dot{a}^2}} \, \Bigg) \,,
     \label{sigma'WH}
\end{eqnarray}
which evaluated at a static solution, $a_0$, shall play a
fundamental role in determining the stability regions. (A
linearized stability analysis of spherically symmetric thin shells
and of thin-shell wormholes where the flux term is zero was
studied in Refs. \cite{linear-WH}).

The surface mass of the thin shell is given $m_s=4\pi a^2 \sigma$.
By rearranging Eq. (\ref{surfenergy}), evaluated at a static
solution $a_0$, one obtains the total mass of the dark energy
star, given by
\begin{equation}\label{totalmass}
M=m(a_0)+m_s(a_0)\left[\sqrt{1-\frac{2m(a_0)}{a_0}}-\frac{m_s(a_0)}{2a_0}\right]
\,.
\end{equation}

Using $m_s=4\pi a^2 \sigma$, and taking into account the radial
derivative of $\sigma'$, Eq. (\ref{consequation2}) can be
rearranged to provide the following relationship
\begin{equation}
\left(\frac{m_s}{2a}\right)''= \Upsilon -4\pi \sigma'\eta \,,
     \label{cons-equation2}
\end{equation}
with the parameter $\eta$ defined as $\eta={\cal P}'/\sigma'$, and
$\Upsilon $ given by
\begin{equation}
\Upsilon \equiv \frac{4\pi}{a}\,(\sigma+{\cal P})+2\pi a \, \Xi '
\,.
\end{equation}
Equation (\ref{cons-equation2}) will play a fundamental role in
determining the stability regions of the respective solutions.
$\eta$ is used as a parametrization of the stable equilibrium, so
that there is no need to specify a surface equation of state. The
parameter $\sqrt{\eta}$ is normally interpreted as the speed of
sound, so that one would expect that $0<\eta \leq 1$, based on the
requirement that the speed of sound should not exceed the speed of
light. We refer the reader to Refs. \cite{linear-WH} for
discussions on the respective physical interpretation of $\eta$ in
the presence of exotic matter.

It is also of interest to analyze the evolution identity, given
by: $\left[\,T_{\mu\nu}\,n^{\mu}n^{\nu}
\right]^+_-=\overline{K}^{i}_{\;\,j}\,S^{j}_{\;\,i}$, where
$\overline{K}^{i}_{\;\,j}=(K^{i\;+}_{\;\,j} +
K^{i\;-}_{\;\,j})/2$. The evolution identity provides the
following relationship
\begin{eqnarray}\label{evolution-identity}
&&p_r+\frac{(\rho+p_r)\dot{a}^2}{1-2m/a}=
-\frac{1}{a}\,\Bigg(\sqrt{1-\frac{2M}{a}+\dot{a}^2}
    \nonumber    \\
&& +\sqrt{1-\frac{2m}{a}+\dot{a}^2}\;\Bigg)\,{\cal P}
     +\frac{1}{2}\Bigg(\frac{M/a^2+\ddot{a}}
{\sqrt{1-\frac{2M}{a}+\dot{a}^2}}
      \nonumber    \\
&& +\frac{m+\omega a
m'+\ddot{a}a^2+\frac{(1+\omega)am'\dot{a}^2}{1-2m/a}}{a^2\sqrt{1-\frac{2m}{a}+\dot{a}^2}}
\Bigg) \sigma \,.
\end{eqnarray}
It is of particular interest to obtain an equation governing the
behavior of the radial pressure in terms of the surface stresses
at the junction boundary, at the static solution $a_0$, with
$\dot{a}=\ddot{a}=0$. From Eq. (\ref{evolution-identity}), we have
the following pressure balance equation
\begin{eqnarray}\label{pressurebalance}
p_r(a_0)&=&
      -\frac{1}{a_0}\,\left(\sqrt{1-\frac{2M}{a_0}}
+\sqrt{1-\frac{2m}{a_0}}\;\right)\,{\cal P}
      \nonumber       \\
&&+\frac{1}{2a_0^2}\left(\frac{M}
{\sqrt{1-\frac{2M}{a_0}}}+\frac{m+\omega a_0
m'}{\sqrt{1-\frac{2m}{a_0}}} \right) \sigma \,.
\end{eqnarray}
Equation (\ref{pressurebalance}) relates the interior radial
pressure impinging on the shell in terms of a combination of the
surface stresses, $\sigma$ and ${\cal P}$, given by eqs.
(\ref{surfenergy})-(\ref{surfpressure}) evaluated at the static
solution, and the geometrical quantities. To gain some insight
into the analysis, consider a zero surface energy density,
$\sigma=0$. Thus, Eq. (\ref{pressurebalance}) reduces to
\begin{equation}
p_r(a_0)=-\frac{2}{a_0}\,\sqrt{1-\frac{2M}{a_0}}\;\,{\cal P} \,.
\end{equation}
As the pressure acting on the shell from the interior is negative
$p_r(a_0)<0$, i.e., a radial tension, then a positive tangential
surface pressure, ${\cal P}>0$, is needed to hold the thin shell
against collapse.

\subsection{Derivation of the master equation}

Equation (\ref{surfenergy}) may be rearranged to provide the thin
shell's equation of motion, i.e., $\dot{a}^2 + V(a)=0$, with the
potential given by
\begin{equation}
V(a)=F(a)-\left[\frac{m_s(a)}{2a}\right]^2-\left[\frac{aG(a)}{m_s(a)}\right]^2
\,.
\end{equation}
where, for notational convenience, the factors $F(a)$ and $G(a)$
are defined as
\begin{eqnarray}
F(a)=1-\frac{m(a)+M}{a} \quad {\rm and} \quad
G(a)=\frac{M-m(a)}{a} \,.
         \label{factor}
\end{eqnarray}

Linearizing around a stable solution situated at $a_0$, we
consider a Taylor expansion of $V(a)$ around $a_0$ to second
order, given by
\begin{eqnarray}
V(a)&=&V(a_0)+V'(a_0)(a-a_0)
     \nonumber  \\
&&+\frac{1}{2}V''(a_0)(a-a_0)^2+O\left[(a-a_0)^3\right] \,.
\label{linear-potential}
\end{eqnarray}
Evaluated at the static solution, at $a=a_0$, we verify that
$V(a_0)=0$ and $V'(a_0)=0$. From the condition $V'(a_0)=0$, one
extracts the following useful equilibrium relationship
\begin{eqnarray}
\Gamma\equiv\left(\frac{m_s}{2a_0}\right)'
=\left(\frac{a_0}{m_s}\right)\left[
F'-2\left(\frac{a_0G}{m_s}\right)\left(\frac{a_0G}{m_s}\right)'\right]
  \,,
\end{eqnarray}
which will be used in determining the master equation, responsible
for dictating the stable equilibrium configurations.

The solution is stable if and only if $V(a)$ has a local minimum
at $a_0$ and $V''(a_0)>0$ is verified. Thus, from the latter
stability condition, one may deduce the master equation, given by
\begin{equation}
\eta_0 \, \frac{d\sigma^2}{da}\Big|_{a_0} > \Theta\,,
\end{equation}
by using Eq. (\ref{cons-equation2}), where $\eta_0=\eta(a_0)$ and
$\Theta$ is defined by
\begin{equation}
\Theta \equiv \frac{1}{2\pi}\left[\sigma \Upsilon+\frac{1}{2\pi
a_0} \left(\Gamma^2-\Psi \right) \right] \,,
       \label{master}
\end{equation}
with
\begin{eqnarray}
\Psi=\frac{F''}{2}-\left[\left(\frac{aG}{m_s}\right)'\right]^2
-\left(\frac{aG}{m_s}\right)\left(\frac{aG}{m_s}\right)'' \,.
\end{eqnarray}

Now, from the master equation we find that the stable equilibrium
regions are dictated by the following inequalities
\begin{eqnarray}
\eta_0 &>& \Omega, \qquad {\rm if} \qquad
\frac{d\sigma^2}{da}\Big|_{a_0}>0\,,      \label{stability1}
       \\
\eta_0 &<& \Omega, \qquad {\rm if} \qquad
\frac{d\sigma^2}{da}\Big|_{a_0}<0\,,       \label{stability2}
\end{eqnarray}
with the definition
\begin{eqnarray}
\Omega\equiv
\Theta\left(\frac{d\sigma^2}{da}\Big|_{a_0}\right)^{-1}\,.
\end{eqnarray}

We shall now model the dark energy stars by choosing specific mass
functions, and consequently determine the stability regions
dictated by the inequalities
(\ref{stability1})-(\ref{stability2}). In the specific cases that
follow, the explicit form of $\Omega$ is extremely messy, so that
as in \cite{LoboCrawford,stableWH}, we find it more instructive to
show the stability regions graphically.

\subsection{Stability regions}

There is only the need to specify the mass function $m(r)$, as the
``gravity profile'' $g(r)$ has been eliminated from the stability
analysis in this section, by using Eq. (\ref{EOScondition})
evaluated at $a$. In the examples that follow, we shall adopt a
conservative point of view, by interpreting that $\sqrt{\eta}$ is
the speed of sound, and taking into account the requirement that
the latter should not exceed the speed of light, i.e., $0<\eta
\leq 1$, on the surface layer. We shall also impose a positive
surface energy density, $\sigma>0$, which implies $m(a)<M$. As
mentioned above, we shall not show the specific form of the
functions $\Xi$, $\Theta$ and $\Omega$, leading to the master
equation, as they are extremely lengthy. However, the stability
regions will be shown graphically.

\subsubsection{1. Constant energy density}

Consider the specific case of a constant energy density, with the
mass function and the gravity profile, which we shall include for
self-completeness, given by
\begin{eqnarray}
m(r)&=& Ar^3
       \\
g(r) &=&\frac{Ar(1+\omega)}{1-2Ar^2}
\end{eqnarray}
with $A=4\pi \rho_0/3$. We shall impose a positive surface energy
density, $\sigma>0$, so that $m(a)<M$. Note that this latter
condition, with $a>2M$, places an upper bound on the constant
energy density of the star's interior, namely, $\rho_0<3/(32\pi
M^2)$.

\begin{figure}[h]
\centering
  \includegraphics[width=2.6in]{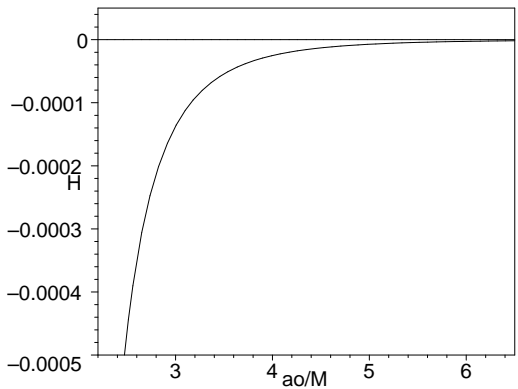}
  \caption{We have defined the dimensionless
parameter $H=M^3 d\sigma^2/da|_{a_0}$. Note that for the case of
$m(a)<M$, we have $H<0$. The latter is shown for the specific case
of $m(a)=M/2$. }
  \label{Fig:equilibrium}
\end{figure}

For the case of $m(a)<M$, one may prove that
$d\sigma^2/da|_{a_0}<0$, so that the stability regions are
dictated by inequality (\ref{stability2}). The latter is shown
graphically in Fig. \ref{Fig:equilibrium}, for the specific case
of $m(a)=M/2$.

\begin{figure}[h]
\centering
  \includegraphics[width=2.8in]{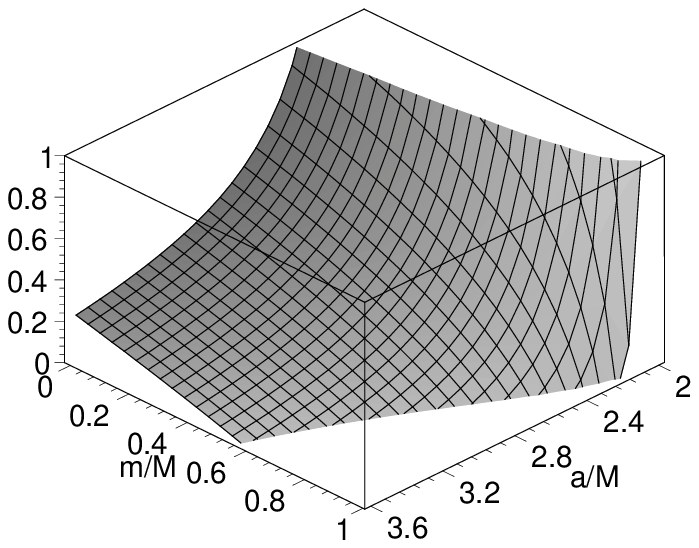}
  \includegraphics[width=2.8in]{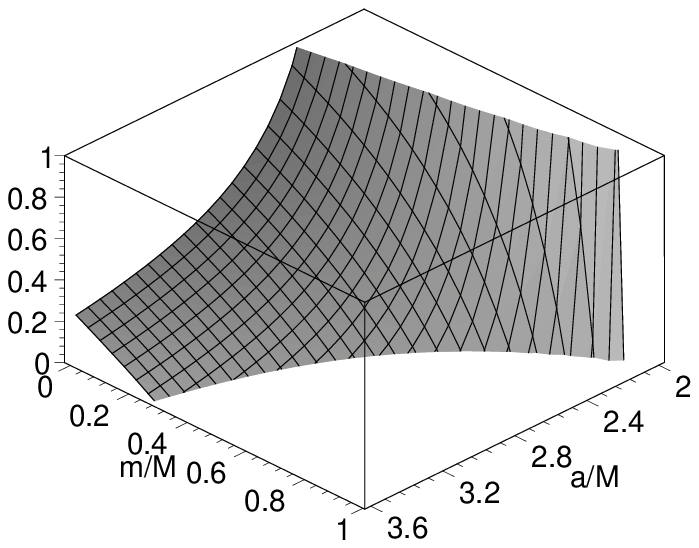}
  \caption{Plots of the stability regions for a dark energy star with
  constant energy
  density. We have considered $\omega=-0.5$ and $\omega=-1.25$,
  in the first and second plots, respectively. The stability regions
  are given below the surfaces. Qualitatively, one verifies that the
  stability regions decrease for a decreasing dark energy parameter.}
  \label{Fig:const-rho}
\end{figure}

Considering the cases of $\omega=-0.5$ and $\omega=-1.25$, the
stability regions are given by the plots depicted below the
surfaces in Fig. \ref{Fig:const-rho}. Note that the stability
regions are sufficiently close to the event horizon, which is
extremely promising. For this case, the stability regions decrease
for decreasing $\omega$, i.e., as the dark energy parameter drops
into the phantom regime. Note that, qualitatively, as
$m\rightarrow M$, the only stability regions that exist are in the
neighborhood of where the event horizon is expected to form.

The above analysis shows that stable configurations of the surface
layer, located sufficiently near to where the event horizon is
expected to form, do indeed exist. Therefore, considering these
models, one may conclude that the exterior geometry of a dark
energy star would be practically indistinguishable from an
astrophysical black hole.

\subsubsection{2. Tolman-Matese-Whitman mass function}

Consider the Tolman-Matese-Whitman mass function, which represents
a monotonic decreasing energy density in the star interior, and
the respective ``gravity profile'', given by
\begin{eqnarray}
m(r)&=&\frac{b_0 r^3}{2(1+2b_0 r^2)}\,,
       \\
g(r) &=&\left(\frac{b_0r}{2}\right)\left[\frac{(1+3\omega)
+(1+\omega)2b_0r^2}{(1+b_0r^2)(1+2b_0r^2)} \right]\,.
\end{eqnarray}

Note that the constant $b_0$ may be expressed in terms of the mass
function, as $b_0=2m [a^3(1-4m/a)]^{-1}$.
Now, as $b_0$ is considered to be non-negative, by construction,
then an additional restriction needs to be imposed, namely,
$a>4m(a)$. However, as we are primarily interested in the behavior
of the model's surface layer where the event horizon is expected
to form, i.e., $a\gtrsim 2M$, we shall only analyze the domain
$0<m/M \leq 1/2$.

\begin{figure}[h]
\centering
  \includegraphics[width=2.8in]{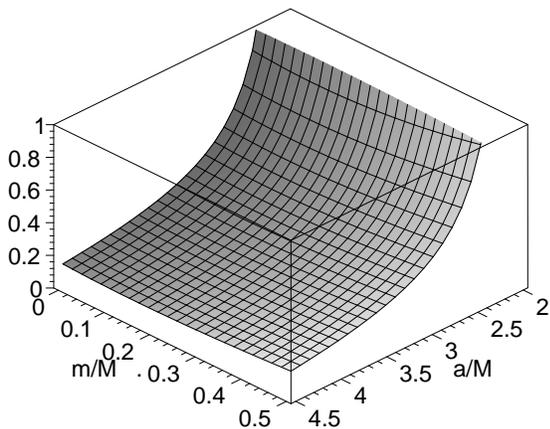}
  \caption{Plot of the stability region for a dark energy star with the
  Tolman-Matese-Whitman mass function, where we have considered
  the case of $\omega=-0.5$.
  The stability region is given below the surface.
  See the text for details.}
  \label{Fig:TMW}
\end{figure}

For the case of $m(a)<M$, as in the previous example, one may also
prove that $d\sigma^2/da|_{a_0}<0$, so that the stability regions
are also dictated by inequality (\ref{stability2}). The stability
regions are depicted in Fig. \ref{Fig:TMW}, for the specific case
of $\omega=-0.5$. It is possible to show that the stability
regions, for this specific case, are very insensitive to
variations in $\omega$. One may prove that the stability
configurations slightly increase for high $m/M$, and slightly
decrease for relatively low values of $m/M$, for decreasing values
of the dark energy parameter $\omega$.

The message that one may extract, as in the previous example, is
that stable dark energy stars exist with a transition layer placed
sufficiently close to where the event horizon is expected to form,
so that the exterior geometry of these stars would be difficult to
distinguish from an astrophysical black hole.

\section{Summary and Conclusion}

Although evidence for the existence of black holes is very
convincing, a certain amount of scepticism regarding the physical
reality of event horizons is still encountered, and it has been
argued that despite the fact that observational data do indeed
provide strong arguments in favor of event horizons, they cannot
fundamentally prove their existence \cite{AKL}. In part, due to
this scepticism, a new picture for an alternative final state of
gravitational collapse has emerged, where an interior compact
object is matched to an exterior Schwarzschild vacuum spacetime,
at or near where the event horizon is expected to form. Therefore,
these alternative models do not possess a singularity at the
origin and have no event horizon, as its rigid surface is located
at a radius slightly greater than the Schwarzschild radius. In
particular, the gravastar picture, proposed by Mazur and Mottola
\cite{Mazur}, has an effective phase transition at/near where the
event horizon is expected to form, and the interior is replaced by
a de Sitter condensate. The latter is then matched to a thick
layer, with an equation of state given by $p=\rho$, which is in
turn matched to an exterior Schwarzschild solution. A simplified
model was then proposed by Visser and Wiltshire \cite{VW}, where
the matching occurred  at a thin shell. It has also been argued
that there is no way of distinguishing a Schwarzschild black hole
from a gravastar from observational data \cite{AKL}.
In this work, a generalization of the gravastar picture was
explored, by considering a matching of an interior solution
governed by the dark energy equation of state, $\omega\equiv p/
\rho<-1/3$, to an exterior Schwarzschild vacuum solution at a
junction interface. We emphasize that the motivation for
implementing this generalization arises from the fact that recent
observations have confirmed an accelerated expansion of the
Universe, for which dark energy, a cosmic fluid dictated by an
equation of state given by $\omega=p/\rho$ with $\omega<-1/3$, is
a possible candidate. We have analyzed several relativistic dark
energy stellar configurations by imposing specific choices for the
mass function. The first case considered was that of a constant
energy density, and the second choice, that of a monotonic
decreasing energy density in the star's interior. We then further
explored the dynamical stability of the transition layer of these
dark energy stars to linearized spherically symmetric radial
perturbations about static equilibrium solutions. It was found
that large stability regions do exist, which are located
sufficiently close to where the event horizon is expected to form,
so that it would be difficult to distinguish the exterior geometry
of the dark energy stars, analyzed in this work, from an
astrophysical black hole.

The possibility of the existence of dark energy, responsible for
the present accelerated expansion of the Universe, has opened up
new possibilities in theoretical research. In this context, by
extending the notion of a spatially homogeneous dark energy fluid,
to inhomogeneous spherically symmetric spacetimes, such as dark
energy stars, we point out that the latter have some interesting
physical properties. Note that recent fits to supernovae, CMB and
weak gravitational lensing data favor an evolving equation of
state, with dark energy parameter crossing the phantom divide
$\omega=-1$ \cite{Vikman,phantom-divide}. Thus, in a rather
speculative scenario, one may consider the existence of a dark
energy star, with an evolving parameter starting out in the range
$-1<\omega<-1/3$, and crossing the phantom divide, $\omega=-1$. A
possible approach would be to consider a mixture of interacting
non-ideal fluids, as it seems that in a cosmological setting the
transition into the phantom regime is physically implausible, for
a single scalar field \cite{Vikman}. One could also apply
variations of the approach outlined in Ref. \cite{Picon-Lim},
where an inhomogeneous spherically symmetric dark energy scalar
field was considered, although the latter model does not allow a
phantom crossing. Once in the phantom regime, the null energy
condition is violated, which physically implies that the negative
radial pressure exceeds the energy density. Therefore, an enormous
pressure in the center may, in principle, imply a topology change,
consequently opening up a tunnel, and converting the dark energy
star into a wormhole \cite{Morris,Visser}. (However, it is still
uncertain whether topology changes will be permitted by an
eventual theory of quantum gravity). It has recently been shown
that traversable wormholes may, in principle, be supported by
phantom energy \cite{Sushkov,Lobo-phantom}, which apart from being
used as interstellar shortcuts, may induce closed timelike curves
with the associated causality violations \cite{mty,Visser}. It
should be interesting to construct a mathematical model
illustrating this conversion, i.e., dark energy star into a
wormhole, by considering a time-dependent dark energy parameter,
which we leave for a future work. Perhaps not so appealing, one
could denote these exotic geometries consisting of dark energy
stars (in the phantom regime) and phantom wormholes as {\it
phantom stars}. As emphasized in the Conclusion of Ref.
\cite{CFV}, we would also like to state our agnostic position
relatively to the existence of dark energy stars, however, it is
important to understand their general properties to further
understand the observational data of astrophysical black holes.

\section*{Acknowledgements}

We thank Alexander Vikman for helpful comments, relatively to the
phantom crossing of inhomogeneous spherically symmetric
distributions of matter.



\end{document}